\documentclass{article}
\usepackage{spconf,amsmath,graphicx}

\usepackage{url}

\usepackage{array,multirow,graphicx}
\usepackage{float}
\usepackage{graphicx}
\usepackage{xparse,xfp}
\usepackage{amsmath, amssymb}
\usepackage{cite}
\usepackage{tikz}
\usepackage{float}
\usepackage{hyperref}
\usepackage[capitalize]{cleveref}
\usetikzlibrary{shapes,arrows, decorations}
\usepackage{pifont}
\usepackage{microtype}      
\usepackage{enumitem}
\usepackage{subcaption}
\hypersetup{pdfauthor={Luca Della Libera, Mirco Ravanelli},
pdftitle={Focal Modulation Networks for Interpretable Sound Classification},
pdfsubject={Interpretable sound classification},
pdfkeywords={Sound classification, focal modulation networks, interpretability, deep learning}}

\newcommand{\luca}
[1]{#1}
\newcommand{\cem}
[1]{#1}

\title{Focal Modulation Networks for Interpretable Sound Classification}

\name{
\parbox{\linewidth}{\centering
{Luca Della Libera$^{1}$, Cem Subakan$^{2,1,3}$, Mirco Ravanelli$^{1,3}$}}
}

\address{
\parbox{\linewidth}{\centering
$^1$Concordia University, $^2$Université Laval, $^3$Mila-Quebec AI Institute}
}

\begin{document}
\ninept

\setlength{\abovedisplayskip}{3.5pt}
\setlength{\belowdisplayskip}{3.5pt}

\maketitle

\begin{abstract}
The increasing success of deep neural networks has raised concerns about their inherent black-box nature, posing challenges related to interpretability and trust. While there has been extensive exploration of interpretation techniques in vision and language, interpretability in the audio domain has received limited attention, primarily focusing on post-hoc explanations.
This paper addresses the problem of interpretability by-design in the audio domain by utilizing the recently proposed attention-free focal modulation networks (FocalNets).
We apply FocalNets to the task of environmental sound classification for the first time and evaluate their interpretability properties on the popular ESC-50 dataset. Our method outperforms a similarly sized vision transformer both in terms of accuracy and interpretability. Furthermore, it is competitive against PIQ, a method specifically designed for post-hoc interpretation in the audio domain.
\end{abstract}


\section{Introduction}
In recent years, deep learning has made remarkable progress in a variety of fields including computer vision\cite{rombach2022high,kirillov2023segany}, natural language processing~\cite{openai2023gpt4,touvron2023llama}, and speech recognition~\cite{radford2022robust,borsos2023audiolm}. Large transformer-based models like SAM~\cite{kirillov2023segany}, GPT-4~\cite{openai2023gpt4}, Whisper~\cite{radford2022robust} have revolutionized their respective domains, often showing superhuman performance.
However, the impressive achievements of deep neural networks are accompanied by a growing concern about their inherent black-box nature. The lack of interpretability in these models, i.e., the ability to understand and explain their predictions, not only challenges people's trust in deep learning systems but also raises ethical issues, including algorithmic discrimination~\cite{zhang2021survey}. This gap poses significant challenges and restricts the applicability of deep learning models, especially in sensitive fields such as autonomous driving, healthcare, criminal justice, and financial services~\cite{li2022interpretability}, where an error may lead to catastrophic consequences. In domains where understanding and trusting AI system decisions are crucial, addressing this aspect becomes fundamental for successful integration in practical use cases.

Several approaches exist to construct interpretable models, which can be broadly categorized as \emph{post-hoc} or \emph{by-design}~\cite{parekh2023tackling}. Post-hoc methods rely on the availability of a predictive model trained and optimized for performance but lacking interpretability. The objective is to implement an auxiliary module that interprets the existing model, a posteriori. Examples for this class include concept-based methods such as TCAV~\cite{kim2018tcav}, ACE~\cite{ghorbani2019}, ConceptSHAP~\cite{yeh2020completeness}, and FLINT~\cite{parekh2020flint}, and attribution-based methods such as integrated gradients~\cite{sundararajan2017ig}, SHAP~\cite{lundberg2017unified},
GradCAM~\cite{selvaraju2016gradcam} and LIME~\cite{ribeiro2016lime}.
Conversely, by-design approaches aim to construct an interpretable predictive model directly from the data. This typically involves constraining the model's capacity, such as by defining prototypes
~\cite{zarei2022prototypical}.
The challenge in this scenario is to achieve optimal interpretability while concurrently preserving performance within the same model. Examples for this class include GAME~\cite{lee2019functional}, INVASE~\cite{yoon2018invase}, FRESH~\cite{jain2020learning}, and SENN~\cite{alvarez2018self}.
While these techniques have been extensively explored in vision and language, interpretability in the audio domain has received limited attention.
Attribution-map based strategies such as LRP~\cite{montavon2018interpreting} and GuidedBackprop~\cite{springenberg2014guidedbackprop} have been applied to audio digit recognition~\cite{becker2018lrp} and feature importance analysis in time-domain CNNs~\cite{muckenhirn2019relevance}, respectively.
Nevertheless, these approaches fall short in adequately addressing the aspect of listenability. 
AudioLIME~\cite{haunschmid2020audiolime} mitigates this issue by separating the input using predefined sources to create simplified audio representations. Although beneficial, it is not ideal as it requires the existence of known and meaningful sources in the input audio.
More sophisticated methods such as FLINT~\cite{parekh2020flint}, L2I~\cite{parekh2022l2i}, and PIQ~\cite{paissan2023piq}, which involve training an additional component -- the interpreter -- are more flexible in this regard, and they can produce more plausible audio interpretations.\looseness-1

All the aforementioned methods primarily focus on post-hoc explanations. On the contrary, when it comes to interpretability by-design in the audio domain, the existing literature is scarce.
A notable example is APNet~\cite{zinemanas2021apnet}, which adapts prototypical networks~\cite{li2018protodnn, chen2019looks}
to audio input by defining a more suitable distance measure for audio prototypes.
Another research effort in this direction is SincNet~\cite{ravanelli2018sincnet}, which encourages the learning of more interpretable filters in the first layer of CNNs by means of parameterized sinc functions.

In this work, we build on the recently proposed focal modulation networks -- FocalNets~\cite{yang2022focalnets} -- to address the challenge of interpretability by-design. FocalNets have demonstrated exceptional interpretability in image classification and segmentation tasks and have proven successful across various applications, including medical imaging~\cite{salari2023focalerrornet}, video action recognition~\cite{wasim2023video}, and forensics~\cite{das2023forensics}.
However, it remains unclear to what extent these interpretability properties hold for the audio domain. To answer this question, we apply FocalNets to the task of environmental sound classification for the first time and we analyze their interpretation capabilities.
In particular, our contributions are as follows:
\begin{itemize}[leftmargin=0.35cm]
    \item We show that FocalNets, which are interpretable by-design, reach a competitive interpretation performance on the ESC-50~\cite{piczak2015esc} dataset.
    \item We introduce a simple strategy to interpret a trained FocalNet classifier, which can be extended with discrete success to transformers.\looseness-1
    \item Our method outperforms a similarly sized vision transformer both in terms of accuracy and interpretability, and is competitive against PIQ~\cite{paissan2023piq}, a method specifically designed for post-hoc interpretation in the audio domain.\looseness-1
\end{itemize}

\section{Method}

\subsection{Focal Modulation Networks}
\begin{figure}[t]
  \centering
  \includegraphics[width=0.36\textwidth]{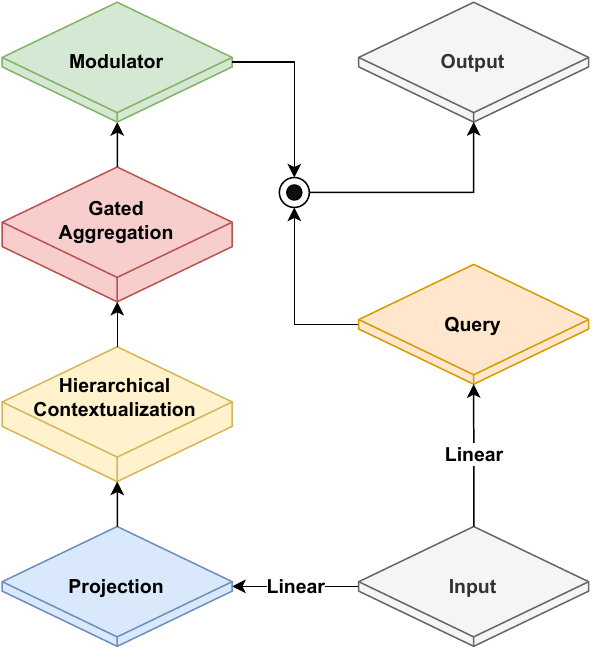}\vspace{-.2cm}
  \caption{The focal modulation layer~\cite{yang2022focalnets}.}
  \label{fig:focal_modulation}
  \vspace{-0.6cm}
\end{figure}

Similarly to the vision transformer (ViT)~\cite{dosovitskiy2021vit}, FocalNets~\cite{yang2022focalnets} were designed to capture contextual information within images. However, while ViT's self-attention produces a feature representation $\mathbf{y}_i \in \mathbb{R}^C$ for each token \luca{(i.e., location)} $\mathbf{x}_i \in \mathbb{R}^C$ in an input feature map $\mathbf{X} \in \mathbb{R}^{C \times H \times W}$ by computing an interaction with its neighbors in $\mathbf{X}$ and aggregating over the contexts, FocalNets' modulation (see \cref{fig:focal_modulation}) first generates the context of the entire input through an aggregation and then computes the modulated interaction with this aggregated vector.
This enables the interactions to be focused on the actual context of the input, as opposed to being influenced by specific values.
The \textbf{focal modulation} value for each query $\mathbf{x}_i \in \mathbb{R}^{C}$ is computed as
\begin{equation}
    \mathbf{y}_i = q(\mathbf{x}_i) \odot m(i, \mathbf{X}),
\end{equation}
where $q(\cdot)$ is a linear transform, $m(\cdot)$ is a context aggregation function whose output is called \emph{modulator}, and $\odot$ is the element-wise multiplication operation.
The modulator is defined as
\begin{equation}
    \label{eq:focal_modulation}
    m(i, \mathbf{X}) = h \left( \sum_{\ell=1}^{L+1} \mathbf{g}^\ell_i \cdot \mathbf{z}^\ell_i \right),
\end{equation}
where $h(\cdot)$ is a linear transform, and $\mathbf{g}^\ell_i \in \mathbf{G}^\ell$ and $\mathbf{z}^\ell_i \in \mathbf{Z}^\ell$ are the gating and context values at location $i$ and focal level $\ell \in \{1, \dots, L + 1\}$, with $\mathbf{G}^\ell \in \mathbb{R}^{1 \times H \times W}$ and $\mathbf{Z}^\ell \in \mathbb{R}^{C \times H \times W}$ denoting the respective \emph{sets} of gating and context values.

The set of context values $\mathbf{Z}^{\ell}$ is obtained via \textbf{hierarchical contextualization}, which consists of a sequence of depth-wise convolutional layers to address long to short-range dependencies.
More specifically, given an input feature map $\mathbf{X} \in \mathbb{R}^{C \times H \times W}$, it is first linearly projected into a feature map $\mathbf{Z}^0 = f_z(\mathbf{X}) \in \mathbb{R}^{C \times H \times W}$. Then, a hierarchical representation of contexts is obtained through $L$ depth-wise convolutions. At focal level $\ell \in \{1, \dots, L\}$, the output $\mathbf{Z}^\ell$ is computed as
\begin{equation}
\label{eq:hierarchical_contextualization}
\mathbf{Z}^{\ell} = \text{GeLU}( \text{DWConv}^\ell(\mathbf{Z}^{\ell-1} )),
\end{equation}
where DWConv$^\ell$ is the depth-wise convolution operator with kernel size $k^\ell$ and GeLU is the Gaussian error linear unit activation~\cite{hendrycks2020gaussian}.
To take into account the global context of the entire input, a global average pooling is applied to the $L$-th focal level feature map to obtain $\mathbf{Z}^{L+1} = \text{AvgPool}(\mathbf{Z}^{L})$. Consequently, a total of $L + 1$ feature maps $\{\mathbf{Z}^{\ell}\}_{\ell=1}^{L+1}$ are generated, collectively capturing contexts at different levels of granularity.

The set of gating values $\mathbf{G}^{\ell}$ is computed during the \textbf{gated aggregation} step, which involves gathering context values from hierarchical contextualization to construct the modulator $m(i, \mathbf{X})$ for each query token.
In practice, a linear transform is used to derive spatial and level-aware gating weights:
\begin{equation}
\label{eq:gating_weights}
    \mathbf{G}^\ell = f_g^l(\mathbf{X}).
\end{equation}
The calculated context values $\mathbf{g}^\ell_i \in \mathbf{G}^\ell$ and gating values $\mathbf{z}^\ell_i \in \mathbf{Z}^\ell$ are then 
aggregated to form $m(i, \mathbf{X})$ according to \cref{eq:focal_modulation}.
This design allows for several properties useful for image processing, such as translation invariance, explicit input-dependency, spatial and channel-specificity, and decoupled feature granularity~\cite{yang2022focalnets}.

\subsection{FocalNet Architecture}
Following the original FocalNet paper~\cite{yang2022focalnets}, we build a focal block by replacing the self-attention mechanism in a standard transformer block~\cite{vaswani2017transformer} with focal modulation. A full backbone can then be constructed via a multi-stage approach as in the focal transformer architecture~\cite{yang2021focal}. Each stage comprises a sequence of focal blocks, followed by a patch embedding layer that downsamples the feature map.
Depending on the problem at hand, a task-specific head might be added on top of the backbone. In our scenario, since we focus on environmental sound classification, we perform average pooling on the contextualized representation extracted by the FocalNet, and we add a linear classification head on top.

\subsection{Preprocessing}
As input to the model, we utilize the log-spectrogram representation, i.e., the logarithm of the magnitude of the STFT of the input waveform. 
\luca{Since the model expects RGB images as input, we stack 3 replicas of the log-spectrogram, resulting in the required 3 channels.}
This allows us to treat the audio signal as a natural image and utilize the FocalNet architecture without any modifications.

\subsection{Generating Interpretations}
\label{subsec:generating_interpretations}
As shown in \cite{yang2022focalnets}, FocalNets exhibits strong interpretability, with modulation maps -- i.e., the L2 norm of the modulator across the channel dimension -- gradually focusing on the regions of interest, without explicit guidance.
In particular, the last layer's modulation map $M = \sqrt{\sum_{k=1}^C m_k(i, \mathbf{X})^2}$, where $m_k(i, \mathbf{X})$ is the $k$-th channel of $m(i, \mathbf{X})$, can precisely locate foreground objects.
Based on this observation, we generate an interpretation mask $M^\text{int}$ by thresholding $M$ as follows:
\begin{equation}
    M^\text{int} = \begin{cases}
    1, & \text{if $M \ge M_q$} \\
    0, & \text{otherwise}
  \end{cases},
\end{equation}
where $M_q$ is the $q$-quantile (or quantile of order $q$) of $M$, with $q \in [0, 1]$. Essentially, we interpret the modulation map as a measure of the importance of each input location to the prediction, and we retain only the most relevant regions. Notably, a larger value of $q$ results in a more focused interpretation mask. The final interpretation is then obtained via element-wise multiplication of the input log-spectrogram with the interpretation mask.

\section{Experimental Setup}

\subsection{Dataset} 
We evaluate the proposed approach on the popular ESC-50~\cite{piczak2015esc} dataset, which is as standard benchmark for environmental sound classification tasks. This multi-class dataset comprises 2000 audio recordings sampled at 44.1 kHz, each belonging to one of 50 different environmental sounds, \luca{that include animal sounds, natural soundscapes/water sounds, human/non-speech sounds, interior/domestic sounds, and exterior/urban noises.}
Each audio clip is 5 seconds in duration and is extracted from publicly available recordings on the {\footnotesize\texttt{freesound.org}} project. The dataset is split into five folds. We use the first, second and third fold for training, the fourth for validation, and the fifth for testing.

\subsection{Compared Methods}
We compare the following interpretation methods:
\begin{itemize}[leftmargin=0.35cm]
    \item \textbf{FocalNet (ours)}: we train a FocalNet classifier and generate interpretations as described in \cref{subsec:generating_interpretations}, with a quantile order $q$ set to 0.9. For the backbone, we use the \texttt{focalnet-base}\footnote{\href{https://huggingface.co/microsoft/focalnet-base}{https://huggingface.co/microsoft/focalnet-base}} variant, pretrained on ImageNet-1k~\cite{deng2009imagenet}, which comprises 4 stages with (2, 2, 18, 2) focal blocks, respectively, 2 focal levels with kernel sizes (3, 5), respectively, 128 channels ($C$), a focal window ($L$) of 3, and a patch size of 4. The total number of parameters, including the classification head, is 87.3M.
    
    \item \textbf{ViT}: similarly to FocalNet, we train a ViT classifier and generate interpretations by adapting the procedure described in \cref{subsec:generating_interpretations}, \luca{with a quantile order $q$ set to 0.9 (as in FocalNet)}. Specifically, instead of computing the L2 norm over channels of the modulation map, we compute the average attention map over heads.
    For the backbone, we use the \texttt{vit-base-patch16-224}\footnote{\href{https://huggingface.co/google/vit-base-patch16-224}{https://huggingface.co/google/vit-base-patch16-224}} variant, pretrained on ImageNet-22k~\cite{deng2009imagenet}, which comprises 12 transformer~\cite{vaswani2017transformer} blocks, a hidden size of 768, 12 attention heads, and a patch size of 16. The total number of parameters, including the classification head, is 86.5M.
    
    \item \textbf{FocalNet-PIQ}: we generate interpretations from our trained FocalNet classifier using the recently proposed PIQ~\cite{paissan2023piq} post-hoc interpretation method, which has shown state-of-the-art interpretability performance across several image and audio benchmarks. PIQ trains an interpreter, which consists of an \emph{adapter} and a \emph{decoder}, to learn class-specific latent representations via vector quantization~\cite{vandenord2017vqvae}. This discretization process acts as a bottleneck, directing the interpreter's attention towards the input components crucial for the classifier's decision.
    The interpretation mask, to be multiplied element-wise with the input log-spectrogram, is obtained by using PIQ in binary-masking mode and applying a sigmoid nonlinearity to the interpreter’s output.
    For the interpreter architecture, we follow the original paper~\cite{paissan2023piq}. In particular, we employ a stack of strided residual convolutional layers for the adapter, and a stack of strided transposed convolutional layers for the decoder. Between consecutive layers, we include ReLU activations and batch normalization layers. The backbone's output serves as input for the interpreter.
    The total number of parameters for the interpreter is 71.4M.\looseness-1

    \item \textbf{ViT-PIQ}: following the same procedure as FocalNet-PIQ, we use PIQ to generate post-hoc interpretations from our trained ViT classifier.
    For the interpreter, we use the same architecture described above. Due to the different shape of the classifier latent representations, this results in a total number of parameters for the interpreter of 44.4M.

\end{itemize}

\subsection{Evaluation Metrics}
In addition to \textbf{accuracy (ACC)}, which is used to assess classifier performance, two quantitative metrics are employed to evaluate the generated interpretations. The first one, \textbf{fidelity-to-input (FID-I)}~\cite{paissan2023piq}, measures the percentage agreement between the classifier's predictions for the original input and the generated interpretation. Mathematically, FID-I is expressed as
\begin{equation}
    \text{FID-I} = \frac{1}{N} \sum_{n=1}^N \left [\arg \max_c f_c(x_n) = \arg \max_c f_c(x_n^{\text{int}}) \right ],
\end{equation}
where, $f_c(\cdot)$ is the classifier's output probability for class $c$, $[\cdot]$ is the Iverson bracket which is 1 if the statement is true and 0 otherwise, $x_n$ is the $n$-th data sample, and $x_n^{\text{int}}$ is the corresponding interpretation. FID-I aims to measure how closely the generated interpretations align with the original input in terms of the class predicted by the classifier (larger is better). Ideally, the produced interpretation should not alter the original classifier's decision, ensuring consistency in classification.\looseness-1

The second metric, \textbf{faithfulness (FA)}~\cite{parekh2022l2i}, measures the importance of the interpretation to the classifier decision. Mathematically, FA is expressed as
\begin{equation}
    \label{eq:fa}
    \text{FA} = f_{\widehat c}(x) - f_{\widehat c}(x - x^\text{int}),
\end{equation}
where $f_{\widehat c}(x)$ is the output probability for the class corresponding to the classifier decision $\widehat c$. FA assesses the impact of the interpretation on the classifier decision, emphasizing its relevance in understanding the model's predictions (larger is better).

\subsection{Training Details}
The 5 seconds long audio recordings are first downsampled from 44.1 kHz to 16 kHz.
Then, \luca{following the original PIQ setup}~\cite{paissan2023piq}, log-spectrograms are computed using a 1024-point STFT, with a window length of 23 ms and a hop length of 11 ms. This results in log-spectrograms of size \luca{513 \text{$\times$} 431} \cem{(513 frequency bins and 431 time points)}. Since the pretrained models operate at a resolution of 224 \text{$\times$} 224, we shrink the log-spectrogram to the required size via bilinear interpolation.

The FocalNet and ViT classifiers are trained with a batch size of 16 for 100 epochs using the Adam optimizer~\cite{kingma2015adam} with a cyclical learning rate schedule~\cite{smith2017cyclic} alternating between 1e\scalebox{0.75}[1.0]{$-$}8 and 2e\scalebox{0.75}[1.0]{$-$}4 with a step size of 65000. The weight decay is set 2e\scalebox{0.75}[1.0]{$-$}6. Additionally, we apply gradient clipping to limit the L2 norm of the gradients to 5.
We augment the data by randomly dropping frequency bands and/or chunks with a probability of 75\%.
The models are trained via additive margin softmax loss~\cite{wang2018ams} with a margin of 0.2 and a scale of 30.

\luca{We use the same configuration to train the PIQ interpreters, but we reduce the batch size to 6 due to memory constraints, we use a fixed learning rate of 2e\scalebox{0.75}[1.0]{$-$}4, and we do not employ any data augmentation technique.}
We set the size of the vector quantization dictionary to 1024, and we train the models via the VQ-VAE~\cite{vandenord2017vqvae} training loss with a negative Bernoulli likelihood.

Software for the experimental evaluation was implemented in Python using the SpeechBrain~\cite{ravanelli2021speechbrain} toolkit and is publicly available in the project repository~\footnote{\href{https://github.com/speechbrain/speechbrain/tree/develop/recipes/ESC50}{https://github.com/speechbrain/speechbrain/tree/develop/recipes/ESC50}}.
All the experiments were run on a CentOS Linux machine with an Intel(R) Xeon(R) Silver 4216 Cascade Lake CPU with 32 cores @ 2.10 GHz, 64 GB RAM and an NVIDIA Tesla V100 SXM2 @ 32 GB with CUDA Toolkit 11.7.

\begin{figure*}[t]
    \centering
    \begin{subfigure}{.225\textwidth}
      \includegraphics[width=1.0\linewidth, height=1.0\linewidth]{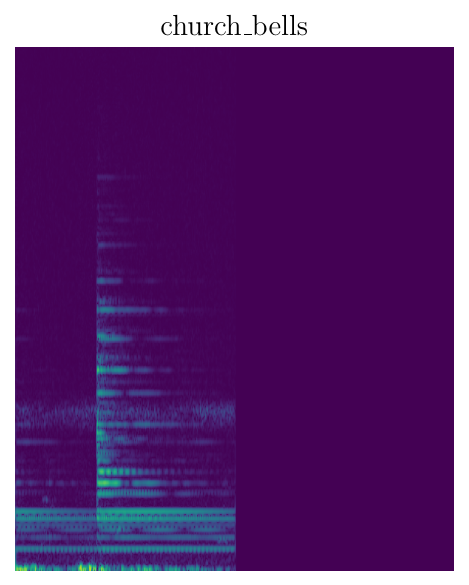}
    \end{subfigure}
    \begin{subfigure}{.225\textwidth}
      \includegraphics[width=1.0\linewidth, height=1.0\linewidth]{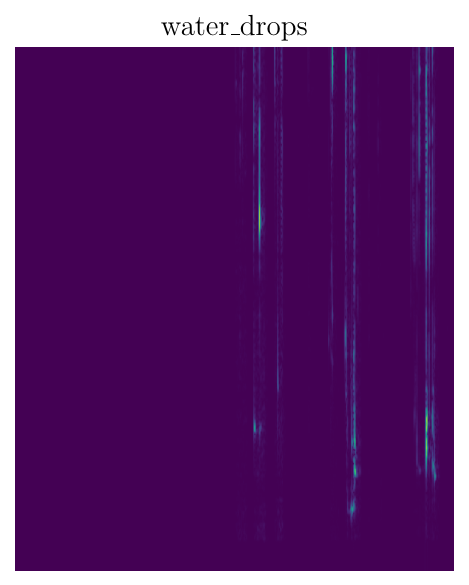}
    \end{subfigure}
    \begin{subfigure}{.225\textwidth}
      \includegraphics[width=1.0\linewidth, height=1.0\linewidth]{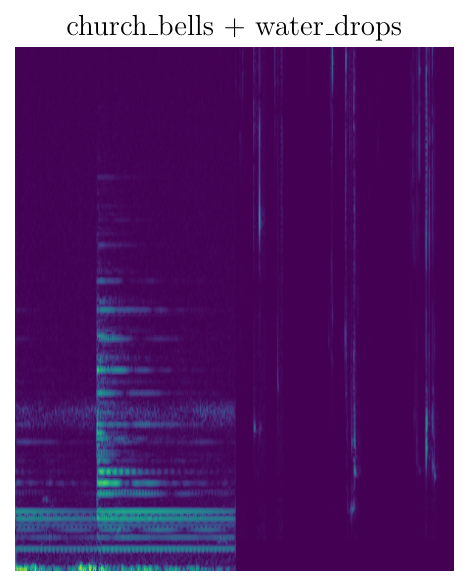}
    \end{subfigure}
    \begin{subfigure}{.225\textwidth}
      \includegraphics[width=1.0\linewidth, height=1.0\linewidth]{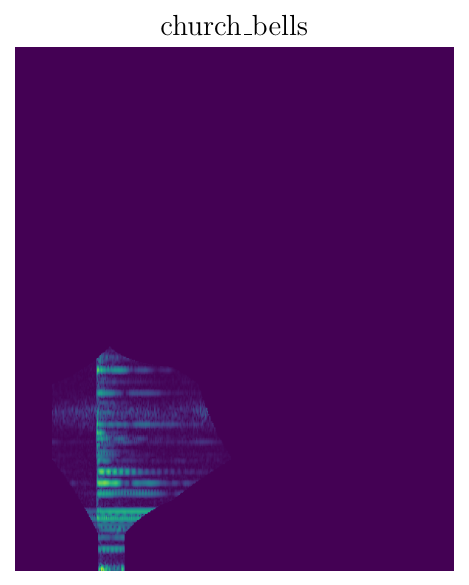}
    \end{subfigure}    
    \vspace{-.25cm}
\caption{
From left to right, the first source ({\footnotesize\texttt{church\_bells}}), the second source ({\footnotesize\texttt{water\_drops}}), the mixture ({\footnotesize\texttt{church\_bells} + \texttt{water\_drops}}), and the interpretation corresponding to the FocalNet's prediction ({\footnotesize\texttt{church\_bells}}). The x-axis represents time, the y-axis represents frequency.
}
\vspace{-0.4cm}
\label{fig:interpretations}
\end{figure*}

\section{Results}

\subsection{Comparative Study}
\label{subsec:comparative_study}
\cref{table:quantitative_results} shows the results of our comparative study. \luca{For quantile order $q$ = 0.9}, FocalNet outperforms all other methods with respect to accuracy and fidelity-to-input, while being competitive in faithfulness.
In particular, we observe that FocalNet is more interpretable than ViT. This is in line with expectations, as attention maps often capture spurious correlations and generally lack interpretability~\cite{wiegreffe2019attention}. Remarkably, FocalNet also achieves better accuracy than ViT, despite being pretrained only on a small subset of the pretraining data used for ViT.

When compared to its post-hoc counterpart, FocalNet demonstrates comparable or even better performance. While post-hoc methods such as PIQ are specifically designed for interpretability, it is encouraging that FocalNet, which aims to maximize accuracy rather then fidelity-to-input or faithfulness, performs well in this regard. In cases where overall performance is similar, by-design approaches like FocalNet are preferred due to their efficiency, as they do not require additional training nor computationally demanding post-processing steps.

A visual example of FocalNet's interpretation on audio is presented in \cref{fig:interpretations}. The input signal comprises {\footnotesize\texttt{church\_bells}} and {\footnotesize\texttt{water\_drops}}, with the first half of the spectrum dominated by {\footnotesize\texttt{church\_bells}} and the second by {\footnotesize\texttt{water\_drops}}. The classifier's prediction is {\footnotesize\texttt{church\_bells}}. It is evident that our method successfully focuses on the relevant portion of the spectrum, emphasizing frequencies contributing more to the sound of {\footnotesize\texttt{church\_bells}}. Importantly, as in PIQ, the generated interpretation can be listened to by reconstructing the waveform via the inverse STFT of the interpretation log-spectrogram.

\begin{table}[t]
\centering
\caption{Evaluation results on ESC-50 test set (fold \#5).}
\vspace{-0.2cm}
\label{table:quantitative_results}
\begin{tabular}{l|c|c|c|c}
\textbf{Method} & \textbf{ACC} ($\uparrow$)  & \textbf{FID-I} ($\uparrow$) & \textbf{FA} ($\uparrow$) & \textbf{\#Params} \\
\hline  
\hline
FocalNet (ours) & \textbf{0.774} & \textbf{0.305} & 0.0111 & 87.3M \\ \hline
ViT & 0.736 & 0.225 & 0.0109 & 86.5M \\ \hline
FocalNet-PIQ & \textbf{0.774} & 0.278 & 0.0111 & 158.7M \\ \hline
ViT-PIQ & 0.736 & 0.110 & \textbf{0.0121} & 130.9M \\ \hline
\end{tabular}
\vspace{-0.6cm}
\end{table}

\subsection{Effect of the Quantile Order}
\begin{figure}[t]
    \centering
    \begin{subfigure}{.238\textwidth}
      \includegraphics[width=1.0\linewidth]{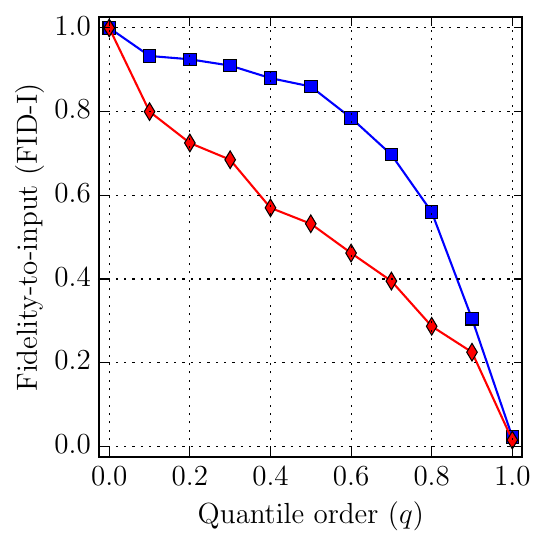}
    \end{subfigure}
    \begin{subfigure}{.238\textwidth}
      \includegraphics[width=1.0\linewidth]{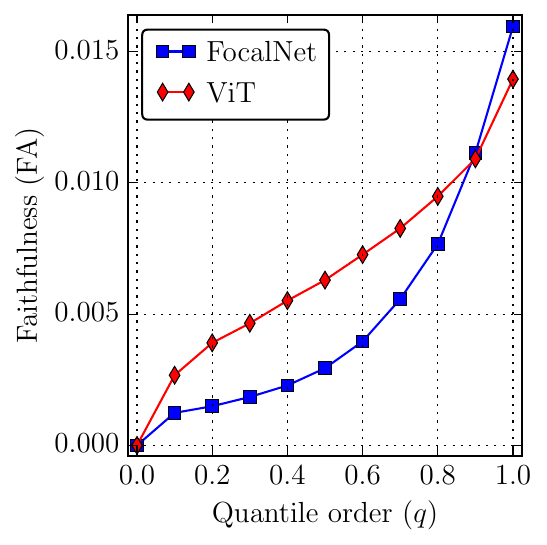}
    \end{subfigure}
    \vspace{-.65cm}
\caption{
The effect of quantile order $q$ on interpretability.
}
\vspace{-0.65cm}
\label{fig:quantile_order}
\end{figure}

\cref{fig:quantile_order} shows the effect of quantile order $q$ on interpretability. We observe that varying $q$ results in a different trade-off between fidelity-to-input and faithfulness. Specifically, when $q$ is close to 1, fidelity-to-input is minimized while faithfulness is maximized. Conversely, when $q$ approaches 0, fidelity-to-input is maximized while faithfulness is minimized.
For larger values of $q$, the interpretation focuses on a small region of the spectrum crucial for prediction. Removing this region would reduce the classifier's confidence for the predicted class, resulting in an increase in faithfulness. \luca{On the contrary}, keeping only this region while masking out the rest of the spectrum may lead to a lack of context, causing the predicted class to change. Thus, a trade-off needs to be found, balancing the removal of noise without sacrificing too much context. This trade-off varies depending on the model. \luca{Note however that, for $q \ge 0.9$, FocalNet outperforms ViT with respect to both fidelity-to-input and faithfulness, while the reverse scenario never occurs, which further confirms our findings from \cref{subsec:comparative_study}.}

\section{Conclusions}
In this paper, we investigated the application of focal modulation networks to the task of interpretable environmental sound classification. 
Our experiments on the ESC-50~\cite{piczak2015esc} dataset show that FocalNet outperforms a vision transformer of similar size, both in terms of accuracy, fidelity-to-input, and faithfulness. Notably, our approach is interpretable by-design, requiring no additional training. Furthermore, it is competitive against post-hoc interpretability methods such as PIQ~\cite{paissan2023piq}. In summary, our findings demonstrate that FocalNets, initially proposed for vision tasks, are also effective in generating interpretations in the audio domain.
\luca{Next steps in this direction include extending the experimental evaluation to other datasets, conducting subjective assessments of the generated listenable interpretations through user studies, and performing sanity checks on the saliency maps using randomization techniques~\cite{adebayo2018sanity}.}

\bibliographystyle{IEEEbib-abbr}
\bibliography{refs}

\begin{thebibliography}{10}

\bibitem{rombach2022high}
R. Rombach, A. Blattmann, D. Lorenz, P. Esser, and B. Ommer,
\newblock ``High-resolution image synthesis with latent diffusion models,''
\newblock in {\em CVPR}, 2022, pp. 10684--10695.

\bibitem{kirillov2023segany}
A. Kirillov et~al.,
\newblock ``Segment anything,''
\newblock {\em arXiv preprint arXiv:2304.02643}, 2023.

\bibitem{openai2023gpt4}
J. Achiam et~al.,
\newblock ``{GPT-4} technical report,''
\newblock {\em arXiv preprint arXiv:2303.08774}, 2023.

\bibitem{touvron2023llama}
H. Touvron et~al.,
\newblock ``{LLaMA}: Open and efficient foundation language models,''
\newblock {\em arXiv preprint arXiv:2302.13971}, 2023.

\bibitem{radford2022robust}
A. Radford et~al.,
\newblock ``Robust speech recognition via large-scale weak supervision,''
\newblock {\em arXiv preprint arXiv:2212.04356}, 2022.

\bibitem{borsos2023audiolm}
Z. Borsos et~al.,
\newblock ``{AudioLM}: a language modeling approach to audio generation,''
\newblock {\em arXiv preprint arXiv:2209.03143}, 2023.

\bibitem{zhang2021survey}
Y. Zhang, P. Tiňo, A. Leonardis, and K. Tang,
\newblock ``A survey on neural network interpretability,''
\newblock {\em IEEE Transactions on Emerging Topics in Computational Intelligence}, pp. 726--742, 2021.

\bibitem{li2022interpretability}
X. Li et~al.,
\newblock ``Interpretable deep learning: Interpretation, interpretability, trustworthiness, and beyond,''
\newblock {\em Knowl. Inf. Syst.}, pp. 3197--3234, 2022.

\bibitem{parekh2023tackling}
J. Parekh, S. Parekh, P. Mozharovskyi, G. Richard, and F. d'Alché Buc,
\newblock ``Tackling interpretability in audio classification networks with non-negative matrix factorization,''
\newblock {\em arXiv preprint arXiv:2305.07132}, 2023.

\bibitem{kim2018tcav}
B. Kim et~al.,
\newblock ``Interpretability beyond feature attribution: Quantitative testing with concept activation vectors ({TCAV}),''
\newblock in {\em ICML}, 2018, pp. 2668--2677.

\bibitem{ghorbani2019}
A. Ghorbani, J. Wexler, J.~Y. Zou, and B. Kim,
\newblock ``Towards automatic concept-based explanations,''
\newblock in {\em NeurIPS}, 2019.

\bibitem{yeh2020completeness}
C.-K. Yeh et~al.,
\newblock ``On completeness-aware concept-based explanations in deep neural networks,''
\newblock in {\em NeurIPS}, 2020.

\bibitem{parekh2020flint}
J. Parekh, P. Mozharovskyi, and F. d'Alch{\'e} Buc,
\newblock ``A framework to learn with interpretation,''
\newblock in {\em NeurIPS}, 2020.

\bibitem{sundararajan2017ig}
M. Sundararajan, A. Taly, and Q. Yan,
\newblock ``Axiomatic attribution for deep networks,''
\newblock in {\em ICML}, 2017, pp. 3319--3328.

\bibitem{lundberg2017unified}
S.~M. Lundberg and S.-I. Lee,
\newblock ``A unified approach to interpreting model predictions,''
\newblock in {\em NeurIPS}, 2017, pp. 4768--4777.

\bibitem{selvaraju2016gradcam}
R.~R. Selvaraju et~al.,
\newblock ``{Grad-CAM}: Visual explanations from deep networks via gradient-based localization,''
\newblock {\em ICCV}, pp. 618--626, 2017.

\bibitem{ribeiro2016lime}
M.~T. Ribeiro, S. Singh, and C. Guestrin,
\newblock ``"{W}hy should i trust you?": Explaining the predictions of any classifier,''
\newblock in {\em ACM SIGKDD ICKDDM}, 2016, pp. 1135--1144.

\bibitem{zarei2022prototypical}
M.~R. Zarei and M. Komeili,
\newblock ``Interpretable concept-based prototypical networks for few-shot learning,''
\newblock in {\em IEEE ICIP}, 2022, pp. 4078--4082.

\bibitem{lee2019functional}
G.-H. Lee, W. Jin, D. Alvarez-Melis, and T.~S. Jaakkola,
\newblock ``Functional transparency for structured data: a game-theoretic approach,''
\newblock in {\em ICML}, 2019.

\bibitem{yoon2018invase}
J. Yoon, J. Jordon, and M. van~der Schaar,
\newblock ``{INVASE}: Instance-wise variable selection using neural networks,''
\newblock in {\em ICLR}, 2019.

\bibitem{jain2020learning}
S. Jain, S. Wiegreffe, Y. Pinter, and B.~C. Wallace,
\newblock ``{L}earning to faithfully rationalize by construction,''
\newblock in {\em ACL}, 2020, pp. 4459--4473.

\bibitem{alvarez2018self}
D. Alvarez-Melis and T.~S. Jaakkola,
\newblock ``Towards robust interpretability with self-explaining neural networks,''
\newblock in {\em NeurIPS}, 2018, pp. 7786--7795.

\bibitem{montavon2018interpreting}
G. Montavon, W. Samek, and K.-R. Müller,
\newblock ``Methods for interpreting and understanding deep neural networks,''
\newblock {\em Digital Signal Processing}, pp. 1--15, 2018.

\bibitem{springenberg2014guidedbackprop}
J.~T. Springenberg, A. Dosovitskiy, T. Brox, and M. Riedmiller,
\newblock ``Striving for simplicity: The all convolutional net,''
\newblock in {\em ICLR Workshop Track}, 2015.

\bibitem{becker2018lrp}
S. Becker, M. Ackermann, S. Lapuschkin, K.-R. M{\"u}ller, and W. Samek,
\newblock ``Interpreting and explaining deep neural networks for classification of audio signals,''
\newblock {\em arXiv preprint arXiv:1807.03418}, 2018.

\bibitem{muckenhirn2019relevance}
H. Muckenhirn, V. Abrol, M. Magimai-Doss, and S. Marcel,
\newblock ``Understanding and visualizing raw waveform-based {CNN}s,''
\newblock in {\em Interspeech}, 2019, pp. 2345--2349.

\bibitem{haunschmid2020audiolime}
V. Haunschmid, E. Manilow, and G. Widmer,
\newblock ``{audioLIME}: Listenable explanations using source separation,''
\newblock in {\em ECML-PKDD}, 2020.

\bibitem{parekh2022l2i}
J. Parekh, S. Parekh, P. Mozharovskyi, F. d'Alch{\'e} Buc, and G. Richard,
\newblock ``Listen to interpret: Post-hoc interpretability for audio networks with {NMF},''
\newblock in {\em NeurIPS}, 2022.

\bibitem{paissan2023piq}
F. Paissan, C. Subakan, and M. Ravanelli,
\newblock ``Posthoc interpretation via quantization,''
\newblock {\em arXiv preprint arXiv:2303.12659}, 2023.

\bibitem{zinemanas2021apnet}
P. Zinemanas, M. Rocamora, M. Miron, F. Font, and X. Serra,
\newblock ``An interpretable deep learning model for automatic sound classification,''
\newblock {\em Electronics}, p. 850, 2021.

\bibitem{li2018protodnn}
O. Li, H. Liu, C. Chen, and C. Rudin,
\newblock ``Deep learning for case-based reasoning through prototypes: A neural network that explains its predictions,''
\newblock in {\em AAAI}, 2018.

\bibitem{chen2019looks}
C. Chen et~al.,
\newblock ``This looks like that: deep learning for interpretable image recognition,''
\newblock in {\em NeurIPS}, 2019, pp. 8928--8939.

\bibitem{ravanelli2018sincnet}
M. Ravanelli and Y. Bengio,
\newblock ``Interpretable convolutional filters with {SincNet},''
\newblock in {\em NeurIPS IRASL Workshop}, 2018.

\bibitem{yang2022focalnets}
J. Yang, C. Li, X. Dai, and J. Gao,
\newblock ``Focal modulation networks,''
\newblock in {\em NeurIPS}, 2022.

\bibitem{salari2023focalerrornet}
S. Salari, A. Rasoulian, H. Rivaz, and Y. Xiao,
\newblock ``{FocalErrorNet}: Uncertainty-aware focal modulation network for inter-modal registration error estimation in ultrasound-guided neurosurgery,''
\newblock in {\em MICCAI}, 2023, pp. 689--698.

\bibitem{wasim2023video}
S.~T. Wasim et~al.,
\newblock ``{Video-FocalNets}: Spatio-temporal focal modulation for video action recognition,''
\newblock in {\em ICCV}, 2023.

\bibitem{das2023forensics}
S. Das and M.~R. Amin,
\newblock ``Learning interpretable forensic representations via local window modulation,''
\newblock in {\em ICCV}, 2023, pp. 436--447.

\bibitem{piczak2015esc}
K.~J. Piczak,
\newblock ``{ESC}: Dataset for environmental sound classification,''
\newblock in {\em ACM MM}, 2015.

\bibitem{dosovitskiy2021vit}
A. Dosovitskiy et~al.,
\newblock ``An image is worth 16x16 words: Transformers for image recognition at scale,''
\newblock in {\em ICLR}, 2021.

\bibitem{hendrycks2020gaussian}
D. Hendrycks and K. Gimpel,
\newblock ``Gaussian error linear units ({GELUs}),''
\newblock {\em arXiv preprint arXiv:1606.08415}, 2020.

\bibitem{vaswani2017transformer}
A. Vaswani et~al.,
\newblock ``Attention is all you need,''
\newblock in {\em NeurIPS}, 2017, pp. 6000--6010.

\bibitem{yang2021focal}
J. Yang et~al.,
\newblock ``Focal attention for long-range interactions in vision transformers,''
\newblock in {\em NeurIPS}, 2021, pp. 30008--30022.

\bibitem{deng2009imagenet}
J. Deng, W. Dong, R. Socher, L.-J. Li, K. Li, and L. Fei-Fei,
\newblock ``{ImageNet}: A large-scale hierarchical image database,''
\newblock in {\em CVPR}, 2009, pp. 248--255.

\bibitem{vandenord2017vqvae}
A. Van~den Oord, O. Vinyals, and K. Kavukcuoglu,
\newblock ``Neural discrete representation learning,''
\newblock in {\em NeurIPS}, 2017.

\bibitem{kingma2015adam}
D.~P. Kingma and J. Ba,
\newblock ``Adam: {A} method for stochastic optimization,''
\newblock in {\em ICLR}, 2015.

\bibitem{smith2017cyclic}
L.~N. Smith,
\newblock ``Cyclical learning rates for training neural networks,''
\newblock in {\em WACV}, 2017, pp. 464--472.

\bibitem{wang2018ams}
F. Wang, J. Cheng, W. Liu, and H. Liu,
\newblock ``Additive margin softmax for face verification,''
\newblock {\em IEEE Signal Processing Letters}, pp. 926--930, 2018.

\bibitem{ravanelli2021speechbrain}
M. Ravanelli et~al.,
\newblock ``{SpeechBrain}: A general-purpose speech toolkit,''
\newblock {\em arXiv preprint arXiv:2106.04624}, 2021.

\bibitem{wiegreffe2019attention}
S. Wiegreffe and Y. Pinter,
\newblock ``Attention is not explanation,''
\newblock in {\em EMNLP-IJCNLP}, 2019, pp. 11--20.

\bibitem{adebayo2018sanity}
J. Adebayo, J. Gilmer, M. Muelly, I. Goodfellow, M. Hardt, and B. Kim,
\newblock ``Sanity checks for saliency maps,''
\newblock in {\em NeurIPS}, 2018, vol.~31.

\end{thebibliography}

\end{document}